\documentclass[12pt]{article} 

\usepackage{graphicx}

\def\1{{\mathchoice {1\mskip-4mu\mathrm l}
{1\mskip-4mu\mathrm l} {1\mskip-4.5mu\mathrm l}
{1\mskip-5mu\mathrm l}}}
\usepackage{amsmath}
\usepackage{amssymb}

\newcommand{\Z}{\mathbb{Z}}
\newcommand{\e}{\varepsilon}
\newtheorem{defi}{Definition}[section]

\newtheorem{Theorem}{Theorem}[section]

\title{Loss and Recovery of Gibbsianness for XY models in external fields}

\author{
{\normalsize A.C.D. van Enter }\\
{\normalsize Intitute for Mathematics and Computing Science,} \\
{\normalsize University of Groningen},\\
{\normalsize Nijenborgh 9, }\\
{\normalsize 9747AG,Groningen},\\
{\normalsize the Netherlands}\\
\\[3 mm]
{\normalsize W.M. Ruszel}\\
{\normalsize Intitute for Mathematics and Computing Science,} \\
{\normalsize University of Groningen},\\
{\normalsize Nijenborgh 9, }\\
{\normalsize 9747AG,Groningen},\\
{\normalsize the Netherlands}\\
{\normalsize and} \\
{\normalsize Center for Theoretical Physics,} \\
{\normalsize University of Groningen},\\
{\normalsize Nijenborgh 4, }\\
{\normalsize 9747AG,Groningen},\\
{\normalsize the Netherlands}}

\begin{document}
\maketitle

\newpage
\begin{abstract}
We consider planar rotors (XY spins) in $\mathbb{Z}^d$,
starting from an initial Gibbs measure and evolving with infinite-temperature
stochastic (diffusive) dynamics. At intermediate times, if the system
starts at low temperature,
Gibbsianness can be lost. Due to the influence of
the external initial field, Gibbsianness can be recovered after large
finite times. We prove some results supporting this picture.
\end{abstract}

\section{Introduction}

Time evolution of spin systems with different initial Gibbs measures and
different dynamics shows various interesting features. In particular,
in the transient regime, the structure of the evolved measure can have
various properties, which may change in time.
For example, in \cite{vEntFerHolRed02},
\cite{vEntRus07}, \cite{KueOpo07}, \cite{KueRed06} and \cite{DerRoe05}
the question was investigated whether the time-evolved measure is
Gibbsian or not. Results about conservation, loss and recovery of
the Gibbs property could be obtained. Ising spin systems were considered
in \cite{vEntFerHolRed02} and different types of unbounded
spin systems in \cite{DerRoe05} and \cite{KueRed06}.
In \cite{vEntRus07} and \cite{KueOpo07} compact continuous
spin systems are investigated.
In more physical terms, the question is whether one can or cannot associate
an effective temperature ($=$ inverse interaction norm) to the system when it
is in this non-equilibrium situation \cite{OlPe07}.

Variations of both the initial and the dynamical temperature (the temperature
of the Gibbs measure(s) to which the system will converge, which is
a property of the dynamics) have influence on
the existence (or absence) of the quasilocality property of the
time-evolved measure of the system. This quasilocality
property is a necessary (and almost sufficient) condition to have
Gibbsianness \cite{EFS93, Geo88}.

In \cite{vEntRus07} we showed that the time-evolved measure for planar rotors
stays Gibbsian for either short times, starting at arbitrary temperature
and with arbitrary-temperature-dynamics, or for high- or infinite-temperature
dynamics starting from a high- or infinite-temperature initial measure for
all times. Furthermore the absence of the quasilocality property is shown for
intermediate times for systems starting in a low-temperature regime with zero
external field and evolving under infinite-temperature dynamics. The fact that
there exist intermediate times where Gibbsianness is lost for XY spins even
in two dimensions is remarkable, because those systems do not have a
first-order phase transition
due to the Mermin-Wagner theorem. However, it turns out that
conditionings can induce one. To establish the occurrence of
such conditional first-order transitions is
a major step in the proof that a certain measure is not Gibbsian.
\newline
Similar short-time results for more general compact spins can be found
in \cite{KueOpo07}. \newline
These results about compact continuous spins can be seen as intermediate
between those for discrete Ising spins and the results for unbounded
continuous spins. Conservation, loss and recovery results can be
found in \cite{vEntFerHolRed02} for Ising spins and conservation for
short times and loss for larger times for unbounded spins in \cite{KueRed06}.
Conservation for short times for more general dynamics (e.g. Kawasaki) for
discrete spins was proven in \cite{LeNRed02}, and for
unbounded spins with bounded interactions in \cite{DerRoe05}.

\medskip

This paper is a continuation of \cite{vEntRus07}. As in that paper,
we consider XY-spins living on a lattice sites on $\mathbb{Z}^d$ and
evolving with time. The initial Gibbs measure is a nearest
neighbour ferromagnet, but now in a positive external field.
So we start in the regime where there is a unique Gibbs
measure. The system is evolving under
infinite-temperature dynamics. We expect, that just as in the Ising case,
whatever the initial field strength, we have after the short times
when the measure is always Gibbsian, if the initial temperature is low, that
a transition towards a non--Gibbsian regime occurs, and that after another,
longer time, the measure becomes Gibbs again.
We can prove a couple of results which go some way in confirming this
picture.

We prove that when the initial field is small, and $d$ is at least 3,
there exists a time interval, depending on the initial field, during which
the time-evolved measure is non-Gibbsian.
We present a partial result, indicating why we expect the same phenomenon to happen in two dimensions.
Furthermore, we argue that the presence of an external field
is responsible for the reentrance into the Gibbsian regime for larger times,
independently of the initial temperature. We can prove this
for the situation in which the original field is strong enough.

\section{Framework and Result}
Let us introduce some definitions and notations. The state space of one
continuous spin is the circle, $\mathbb{S}^1$. We identify the circle with the
interval $[0,2\pi)$ where $0$ and $2\pi$ are considered to be the same points.
Thus the configuration space $\Omega$ of all spins is isomorphic to
$[0,2\pi)^{\Z^d}$. We endow $\Omega$ with the product topology and natural
product probability measure
$d\nu_0(x) = \bigotimes_{i \in \mathbb{Z}^d} d\nu_0(x_i)$.
In our case we take $d\nu_0(x_i) = \frac{1}{2 \pi} dx_i$.
An interaction $\varphi$ is a collection of $\mathcal{F}_{\Lambda}$-measurable
functions $\varphi_{\Lambda}$ from $([0,2\pi))^{\Lambda}$ to $\mathbb{R}$
where $\Lambda \subset \mathbb{Z}^d$ is finite.
$\mathcal{F}_{\Lambda}$ is the $\sigma$-algebra generated by the canonical
projection on $[0,2\pi)^{\Lambda}$. \newline
The interaction $\varphi$ is said to be of \textbf{finite range} if there
exists a $r > 0$ s.t. $diam(\Lambda) > r$ implies $\varphi_{\Lambda} \equiv 0$
and it is called \textbf{absolutely summable} if for all $i$,
$\sum_{\Lambda \ni i} \parallel \varphi_{\Lambda} \parallel_{\infty} < \infty$. \newline
We call $\nu$ a \textbf{Gibbs measure} associated to a reference measure
$\nu_0$ and interaction $\varphi$ if the series
$H_{\Lambda}^{\varphi}= \underset{\Lambda^{\prime} \cap \Lambda \neq \emptyset} \sum \varphi_{\Lambda^{\prime}} $ converges ($\varphi$ is absolutely summable)
and $\nu$ satisfies the DLR equations for all $i$:

\begin{equation}
d\nu_{\beta}(x_i \mid x_j, j \neq i) = \frac{1}{Z_i} \exp( - \beta H_{i}^{\varphi}(x)) d\nu_0( x_i), \label{Gibbs}
\end{equation}

where $Z_i = \int_0^{2\pi} \exp( - \beta H_{i}^{\varphi}(x)) d\nu_0(x)$ is the partition function and $\beta$ proportional to the inverse temperature. The set of all Gibbs measures associated to $\varphi$ and $\nu_0$ is denoted by $\mathcal{G}(\beta, \varphi, \nu_0)$.

Now, instead of working with Gibbs measures on $[0,2\pi)^{\Z^d}$ we will first
investigate Gibbs measures as space-time measures $Q^{\nu_{\beta}}$ on the
path space $\overset{\sim} \Omega = C(\mathbb{R}_+, [0,2\pi))^{\mathbb{Z}^d}$.
In \cite{Deu87} Deuschel introduced and described infinite-dimen\-sional
diffusions as Gibbs measures on the path space $C([0,1])^{\Z^d}$ when the
initial distribution is Gibbsian. This approach was later generalized by
\cite{CatRoeZes96} who showed that there exists a one-to-one correspondence
between the set of initial Gibbs measures and the set of path-space measures
$Q^{\nu_{\beta}}$.

\bigskip

We consider the process $X=(X_i(t))_{t \geq 0, i \in \Z^d}$ defined by
the following system of stochastic differential equations (SDE)
\begin{eqnarray}
\begin{cases}
& d X_i(t) = d B_i^{\odot}(t) , i \in \mathbb{Z}^d, t > 0 \label{system2-1}\\
& X(0) \sim \nu_{\beta} , t=0
\end{cases}
\end{eqnarray}
for $\nu_{\beta} \in \mathcal{G}(\beta, \overset{\sim} \varphi, \nu_0)$ and
the initial interaction $\overset{\sim} \varphi$ given by
\begin{equation}
\overset{\sim} \varphi_{\Lambda}(x) = - J \underset{i,j \in \Lambda: i \sim j} \sum \cos(x_i-x_j) - h\sum_{i \in \Lambda}\cos(x_i) \label{interaction}
\end{equation}
$J, h$ some non-negative constants and $d\nu_0(x) = \frac{1}{2\pi} dx$.
$\overset{\sim}H$ denotes the initial Hamiltonian associated to
$\overset{\sim}\varphi$ and $(B_i^{\odot}(t))_{i,t}$ is
independent Brownian motion moving on a circle with transition kernel given
(via the Poisson summation formula)
\begin{equation*}
p_t^{\odot}(x_i,y_i) = 1 + 2\cdot \sum_{n \geq 1} e^{-n^2 t} \cos(n\cdot(x_i - y_i))
\end{equation*}
for each $i \in \mathbb{Z}^d$, just as we used in \cite{vEntRus07}. Note also that the eigenvalues of the Laplacian on the circle, which is the generator of the process, are given by $\lbrace n^2, n \geq 1 \rbrace$, see also \cite{Ros97}. We remark that the normalization factor $1/2\pi$ is absorbed into the single-site measure $\nu_0$. \newline
Obviously $\overset{\sim}\varphi$ is of finite range and absolutely summable,
so the associated measure $\nu_{\beta}$ given by $\eqref{Gibbs}$ is Gibbs.

\medskip

For the failure of Gibbsianness we will use the necessary and sufficient condition of finding a point of essential discontinuity of (every version of) the conditional probabilities of $\nu_{\beta}$, i.e. a so-called \textbf{bad configuration}.
It is defined as follows
\begin{defi}
A configuration $\zeta$ is called \textbf{bad} for a probability measure $\mu$ if there exists an $\e > 0$ and $i \in \Z^d$ such that for all $\Lambda$ there exists $\Gamma \supset \Lambda$ and configurations $\xi$, $\eta$ such that
\begin{equation}
|\mu_{\Gamma}(X_i | \zeta_{\Lambda \setminus \lbrace i \rbrace}\eta_{\Gamma \setminus \Lambda}) - \mu_{\Gamma}(X_i | \zeta_{\Lambda \setminus \lbrace i \rbrace}\xi_{\Gamma \setminus \Lambda}) | > \e. \label{badconfig}
\end{equation}
\end{defi}
The measure at time $t$ can be viewed as the restriction of the two-layer system, considered at
at times $0$ and $t$ simultaneously, to the second layer. In order to prove Gibbsianness or non-Gibbsianness we need to study the joint Hamiltonian for a fixed value $y$ at time $t$. \newline

The time-evolved measure is \textbf{Gibbsian} if for every fixed
configuration $y$ the joint measure has no phase transition in a strong sense
(e.g. via Dobrushin uniqueness, or via cluster expansion{/}analyticity
arguments). In that case, an
absolutely summable interaction can be found for which the evolved measure
is a Gibbs measure.
On the other side the measure is \textbf{non-Gibbsian} if there exists a
configuration $y$ which induces a phase transition for the conditioned
double-layer measure at time $0$ which can be detected
via the choice of boundary conditions. In that case no such interaction can
be found, see for example \cite{FerPfi97}.
\newline

The results we want to prove are the following.
\begin{Theorem}
Let $Q^{\nu_{\beta}}$ be the law of the solution $X$ of the planar rotor system $\eqref{system2-1}$ in $\Z^d$, $\nu_{\beta} \in \mathcal{G}(\beta, \overset{\sim}\varphi, \nu_0)$ and $\overset{\sim} \varphi$ given by $\eqref{interaction}$, with $\beta$ the inverse temperature, $J$ some non-negative constant and $h> 0$ the external field, and $d$ at least 3.
Then, for $\beta $ large enough, and $h$ small enough, there is a time
interval $(t_0(h,\beta), t_1(h,\beta))$ such that for all
$t_0(h,\beta) < t < t_1(h,\beta)$ the time-evolved measure
$\nu^t_{\beta}=Q^{\nu_{\beta}}\circ X(t)^{-1}$ is not Gibbs, i.e. there exists
no absolute summable interaction $\varphi^t$ such that $\nu^t_{\beta} \in
\mathcal{G}(\beta, \varphi^t, \nu_0)$.
\end{Theorem}

\begin{Theorem}
For any $h$ chosen such that $\beta h$ is large enough,
compared to $\beta$, there exists a
time $t_2(h)$, such that for all $t \geq t_2(h)$
the time-evolved measure is Gibbs, $\nu^t_{\beta} \in \mathcal{G}(\beta,
\varphi^t, \nu_0)$.
\end{Theorem}
\textbf{Proof of Theorem 2.1:}\\
We consider the double layer system, describing the system at times $0$ and $t$. We can rewrite the transition kernel in Hamiltonian form, and we will call the Hamiltonian for the two-layer system the dynamical Hamiltonian, (as it contains the dynamical kernel). It is formally given by: 
\begin{equation*}
-\textbf{H}^t_{\beta}(x,y) = - \beta \overset{\sim}H(x) + \sum_{i \in \Z^2} \log(p_t^{\odot}(x_i,y_i)),
\end{equation*}
where $x, y \in [0,2\pi)^{\Z^d}$, $p_t^{\odot}(x_i,y_i)$ is the transition
kernel on the circle and $\overset{\sim}H(x)$ is formally given by
\begin{equation*}
-\overset{\sim}H(x) = J \sum_{i \sim k} \cos(x_i-x_k) + h\sum_i \cos(x_i).
\end{equation*}

\textbf{1.} First we want to prove that there exists a time interval where Gibbsianness is lost.
For this we have to find a "\textit{bad configuration}" such that the
conditioned double-layer system has a phase transition at time 0, which
implies $\eqref{badconfig}$ for the time-evolved measure.
We expect this to be possible for each strength of the external field, and in
each dimension at least 2. At present we can perform the programme
only for weak fields, and for dimension at least 3. We also show a
partial result, at least indicating how a conditioning also in d=2
can induce a phase transition.
\medskip

Thus, given $h>0$, we immediately see that the spins from the
initial system prefer to follow the field and point upwards (take the
value $x_i=0$ at each site $i$).
To compensate for that, we will condition the system on the configuration
where all spins point downwards (at time $t$),
i.e. $y^{spec}:=(\pi)_{i \in \mathbb{Z}^d}$. Thus the
spin configuration in which all spins point in the direction opposite to the
initial field will be our `` bad configuration''. We expect that then the
minimal configuration of $-\textbf{H}^t_{\beta}(x,y^{spec})$,
so the ground states of the conditioned system at time 0, will need
to compromise between the original field and the dynamical (conditioning) term.
In the ground state(s) either all spins will
point to the right, possibly with a small correction $\e_t$,
$(\pi/2 - \e_t)_{i \in \Z^2}$ or to the left $(3\pi/2 + \e_t)_{i \in \Z^2}$,
also with a small correction.
$\e_t$ is a function depending on $t$. Finally these two symmetry-related
ground states will yield a phase transition of the "spin-flop" type,
also at low temperatures. It is
important to observe that for this intuition to work, it is essential
that the rotation symmetry of the zero-field situation will not be
restored, due to the appearance of higher order terms from the expansion of
the transition kernel, as we will indicate below.

\medskip

We perform a little analysis for the logarithm of the transition kernel
$p_t^{\odot}$. Let $y^{spec}:=(\pi)_{i \in \mathbb{Z}^d}$.
We want to focus on the first three terms coming from the expansion of
the logarithm.
\begin{eqnarray*}
& & \log\biggl ( 1 + 2\sum_{n \geq 1}e^{-n^2 t}\cos(n(x_i-\pi)) \biggr) = \\
& & -2e^{-t}\cos(x_i) - 2e^{-2t}\cos^2(x_i) - \frac{8}{3}e^{-3t}\cos^3(x_i) + R_t(x_i)
\end{eqnarray*}
where
\begin{equation*}
R_t(x_i) := \biggl[ \sum_{n \geq 1} \frac{(-1)^{n+1}}{n}\biggl ( 2 \sum_{k \geq 1} e^{-k^2 t} \cos(k(x_i-\pi))\biggr)^n \biggr] \1_{\lbrace n \neq 1,2,3 \rbrace \cup \lbrace k \neq 1 \rbrace},
\end{equation*}
is of order $\mathcal{O}_i(e^{-4t})$, for details see the Appendix.
We define $h_t=e^{-t}$.
Note that given $\beta h$, there is a time interval where the effect of the
initial field is essentially compensated by the field induced by the dynamics
(containing the $h_t$.
For large times the initial field term
dominates all the others and the system is expected to exhibit
a ground (or Gibbs) state following this field. For intermediate times the other
terms are important, too.
If we consider a small initial field, it is enough to consider
the second and third order terms which we indicated above.
Those terms create, however, the discrete left-right symmetry
for the ground states which will now prefer to point either to the right or to the left.
\newline

For the moment we forget about the rest term $R_t(x_i)$ and investigate the restricted Hamiltonian $-\textbf{H}_{res3}^t(x,y^{spec})$ which is formally equal to
\begin{equation}
\beta J \sum_{i \sim k} \cos(x_i - x_k) + \beta h\sum_i \cos(x_i) + \sum_i \biggl( -2h_t\cos(x_i) - 2h_t^2\cos^2(x_i) - \frac{8}{3}h^{3}_t\cos^3(x_i) \biggr) \label{Cpotential}.
\end{equation}
To be more precise, the external field including the inverse temperature $\beta h$ will be chosen small enough, and then the inverse temperature $\beta$
large enough. We want first to find the ground states of the restricted Hamiltonian
$\textbf{H}_{res3}^t(x,y^{spec})$ which are points
$x=(x_i)_{i \in \mathbb{Z}^d}$.
It is fairly immediate to see that in the ground states all spins point in
the same direction, so we then only need to minimize the single-site energy
terms. The first-order term more or less compensates the external
field, and the second-order term is maximal when $cos^2(x_i)$ is minimal,
thus when one has the value $\pi/2$ or $3 \pi/2$.
The higher-order terms will only minimally change this picture.

We can define a function $\e_t $ depending on $t$ such that asymptotically
$\beta h=h_t + \e_t$ yields the following unique maxima $(\pi/2 - \e_t,\pi/2 -\e_t)$
and $(3\pi/2 + \e_t,3\pi/2 +\e_t)$. The function $\e_t$ is a correction of the ground states pointing to the left or right. We present a schematic illustration of the two ground states.

\begin{minipage}[hbt]{5cm}
\centering
\includegraphics[width= 4 cm, height= 4 cm]{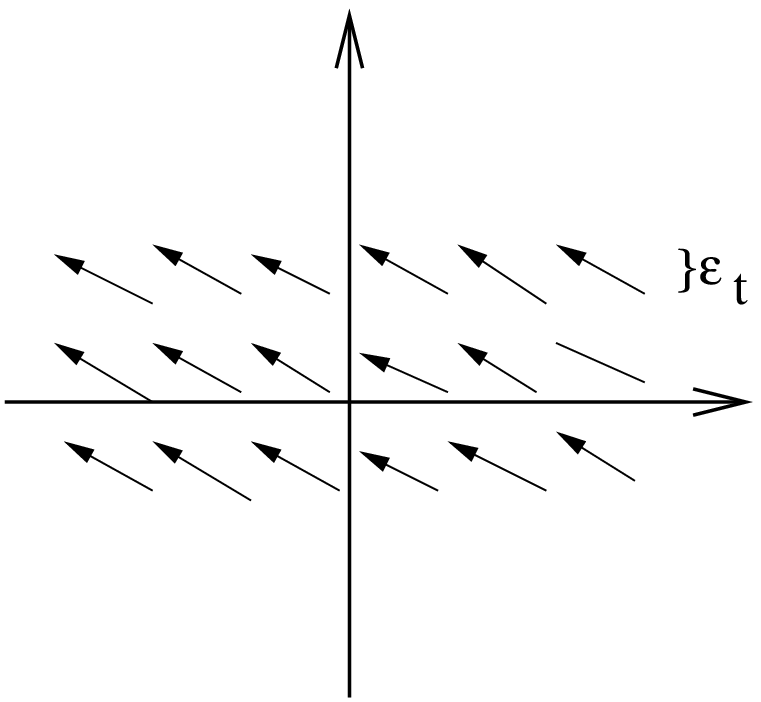}
$(3\pi/2 + \e_t)_i$
\end{minipage}
\hfill
\begin{minipage}[hbt]{5cm}
\centering
\includegraphics[width= 4 cm, height= 4 cm]{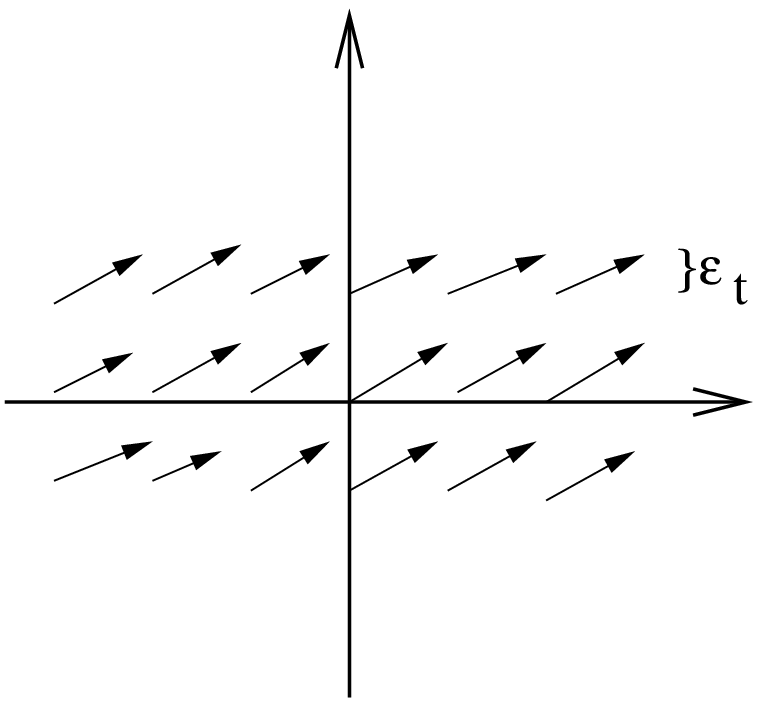}
$(\pi/2 - \e_t)_i$
\end{minipage}

\medskip

Hence for every arbitrarily chosen small external field $h$, we find a time
interval depending on $h$, such that we obtain two reflection
symmetric ground states
of all spins pointing either (almost) to the right
$(\pi/2 -\e_t )_{i\in \Z^d}$ or all spins pointing (almost) to the left
$(3\pi/2 + \e_t)_{i\in \Z^d}$. The rest term $R_t(x_i)$ does not change this
behaviour since it is suppressed by the first terms and is of order
$\mathcal{O}_i(e^{-4t})$. Moreover, it respects the left-right symmetry.

We will first, as a partial argument, show that the interaction
\begin{equation}
 J \sum_{i \sim k} \cos(x_i - x_k) + h\sum_i \cos(x_i) + \sum_i \biggl( -2h_t\cos(x_i) - 2h_t^2\cos^2(x_i) - \frac{8}{3}h^{3}_t\cos^3(x_i) \biggr)
\end{equation}
has a low-temperature transition in $d \geq 2$.

\medskip

To show this we notice that we are in a similar situation as in \cite{vEntRus07}.
The conditioning of the double-layer system for the XY spins created left-right symmetric ground states. 

\medskip

Now we want to apply a percolation argument for low-energy clusters
to prove that such that spontaneous symmetry breaking occurs.
The arguments follow essentially \cite{vEntRus07} and are based on
\cite{Geo81}. The potential corresponding to the Hamiltonian
$\eqref{Cpotential}$ is clearly a $C$-potential, that is a potential
which is nonzero only on subsets of the unit cube \cite{Geo81}.
It is of finite range, translation-invariant and symmetric under
reflections. 

\medskip

Including the rest term (which is a translation-invariant
single-site term) does not change this. A fortiori the
associated measure is reflection positive and we
can again use the same arguments as in \cite{vEntRus07} to deduce that for $\beta$
large enough, there is long-range order. This argument indicates how
conditioning might induce a phase transition.

\bigskip

However, to get back to our original problem, that is, to prove the
non-Gibbsianness of the evolved state
we need an argument which holds for values of not only of $h$, but of
$\beta h$ which are small
uniformly in temperature. Then only we can deduce that
there exists a time interval
$(t_0(\beta,h), t_1(\beta,h))$ such that
$|\mathcal{G}_{\beta}(\textbf{H}^t_{\beta} (\cdot,y^{spec}),\nu_0)| \geq 2$.

To obtain this, for $d=3$, we can invoke a proof using infrared bounds
(see e.g. \cite{FILS78,Geo88,Bis08}. Note that the infrared bound proof, although
primarily developed for proving continuous symmetry breaking, also applies to models with discrete symmetry breaking as we have here. In fact we may include the rest term without any problem here, as the symmetry properties of the complete dynamical Hamiltonian are the same as that of our restricted one, and adding single-site terms does not spoil the reflection positivity.
From this an initial temperature
interval is established, where Gibbsianness is lost after appropriate times.

Indeed, the infrared bound provides a lower bound on the two-point function
which holds uniformly in the single-site measure,
(which in our case varies only slightly anyway,
as long as the field and the compensating
term due to the kernel are small enough). This shows that a phase transition
occurs at sufficiently low temperatures, as for decreasing temperatures the
periodic boundary condition state converges to the symmetric mixture
of the right- and left-pointing ground-state configurations.
\bigskip

\textbf{Comment}: One might expect that, by judiciously looking for other
points of discontinuity, the time interval of proven non-Gibbsianness might be
extended, hopefully also to $d=2$; however, qualitatively this does not
change the picture. In fact,
there are various configurations where one might expect that conditioning on
them will induce a first-order transition. For example, the XY model in at
least two dimensions in a weak random field which is plus or minus with
equal probability is expected to have such transitions \cite{Wehr2006}.
The same situation should occur
for various appropriately chosen (in particular random) choices of
configuration where spins point only up or down.
In a somewhat similar vein, if the original field is not so weak, and thus
also higher terms are non-negligible, we expect that qualitatively
not much changes, and there will again be an intermediate-time regime of
non-Gibbsianness at sufficiently low temperatures.

\bigskip

\textbf{ About the proof of Theorem 2.2:}
Let us now turn to the second statement. Here
the initial temperature does not affect the argument.
The intuitive idea, as mentioned before,
is as follows: As after a long time, the term due to the
conditioning becomes much weaker than the initial external field
-however weak it is-,
uniformly in the conditioning, and thus the system should behave in the
same way as a plane rotor in a homogeneous external field, and have
no phase transition.
However, the higher-order terms which were helpful
for proving the non-Gibbsianness, now prevent us using the ferromagneticity
of the interaction. Indeed, we cannot use correlation
inequalities of FKG type, and we will have to try analyticity methods.

In fact, we expect that the statement should be true for each strength
of the initial field. Indeed, once the time is large enough, the dynamical
single-site term should be dominated by the initial field, and, just as
in that case, one should have no phase transition \cite{Dun79, Dun79a, LieSok81}.
However, to conclude that we can consider the dynamical single-site term as a
small perturbation, in which the free energy and the Gibbs measure are
analytic, although eminently plausible, does not seem to follow from
Dunlop's Yang-Lee theorem.

For high fields, we can either invoke
cluster expansion techniques, showing that the system is Completely Analytic,
or Dobrushin uniqueness statements. Precisely such claims were developed for
proving Gibbsianness of evolved measures at short times
in \cite{vEntRus07} and in \cite{KueOpo07}. A direct application of
those proofs also provides our theorem, which is for long times.

\section{Conclusion}
In this paper we extended the results from \cite{vEntRus07} and show some
results on loss and recovery of Gibbsianness for XY spin systems in an
external field. Giving a low-temperature
initial Gibbs measure in a weak field and evolving with infinite-temperature
dynamics we find a time interval where Gibbsianness is lost. Moreover at large
times and strong initial fields, the evolved measure is a Gibbs measure,
independently of the initial temperature. \newline

Generalizations are possible to include for example more general finite-range,
translation invariant ferromagnetic interactions $\overset{\sim} \varphi$. We
conjecture, but at this point cannot prove,
that both the loss and recovery statements actually hold
for arbitrary strengths of the initial field. \\

\textit{Acknowledgements:} We thank Christof K\"ulske, Alex Opoku, Roberto
Fern\'andez, Cristian Spitoni and especially Frank Redig for helpful
discussions. We thank Roberto Fern\'andez for a careful reading of the
manuscript.
We thank Francois Dunlop for a useful correspondence.

\section{Appendix}

The logarithm of the transition kernel is given by
\begin{equation}
\log\biggl ( 1 + 2\sum_{n \geq 1}e^{-n^2 t}\cos(n(x_i-\pi)) \biggr) = \sum_{k \geq 1} \frac{(-1)^{k+1}}{k}\biggl ( 2 \sum_{n \geq 1} e^{-n^2 t} \cos(n(x_i-\pi))\biggr)^k. \label{logExp}
\end{equation}
Since the first term of the series of $p_t^{\odot}$ is dominating we can write
\begin{equation*}
2\sum_{n \geq 1} e^{-n^2 t} \cos(n(x_i-\pi)) = -2 e^{-t}\cos(x_i) + Rest_t(x_i).
\end{equation*}
The rest term $Rest_t(x_i)$ is smaller than $2e^{-4t}$ uniformly in $x_i$. Then we can bound
\begin{equation*}
2\sum_{n \geq 1} e^{-n^2 t} \cos(n(x_i-\pi)) \leq -2 e^{-t}\cos(x_i) + 2e^{-4t}.
\end{equation*}
Furthermore we write $\eqref{logExp}$ as
\begin{eqnarray*}
& & \biggl ( -2 e^{-t}\cos(x_i) + \mathcal{O}(e^{-4t}) \biggr) - \frac{1}{2}\biggl ( -2 e^{-t}\cos(x_i) + \mathcal{O}(e^{-4t}) \biggr)^2 + \\
& & \frac{1}{3}\biggl ( -2 e^{-t}\cos(x_i) + \mathcal{O}(e^{-4t}) \biggr)^3 +
\sum_{k \geq 4} \frac{(-1)^{k+1}}{k}\biggl ( -2 e^{-t}\cos(x_i) + \mathcal{O}(e^{-4t})\biggr)^k
\end{eqnarray*}
and afterwards bound it by
\begin{equation*}
-2 e^{-t}\cos(x_i) + \mathcal{O}(e^{-4t}) - 2 e^{-2t}\cos^2(x_i) + \mathcal{O}(e^{-5t}) - \frac{8}{3} e^{-3t}\cos^3(x_i) + \mathcal{O}(e^{-6t}) + \mathcal{O}(e^{-4t})
\end{equation*}
thus $\eqref{logExp}$ is then bounded by
\begin{equation*}
-2e^{-t}\cos(x_i) - 2e^{-2t}\cos^2(x_i) - \frac{8}{3}e^{-3t}\cos^3(x_i) + \mathcal{O}(e^{-4t}).
\end{equation*}
Altogether we consider the leading terms of the series $\eqref{logExp}$, $-2e^{-t}\cos(x_i) - 2e^{-2t}\cos^2(x_i) - \frac{8}{3}e^{-3t}\cos^3(x_i)$,
separately and bound the rest uniformly in $x_i$ for every $i$ by $const \times e^{-4t}$ for large $t$.


\begin{thebibliography}{}


\bibitem[${ }^{1}$]{Bis08}
M. Biskup, 
`` Reflection positivity and phase transitions in lattice spin models,''
arXiv: math-ph{/}0610025, Lectures Prague School (2006), to appear.
\\
\bibitem[${ }^{2}$]{CatRoeZes96}
P.Cattiaux, S.Roelly and H. Zessin, 
`` Une approche Gibbsienne des diffusions Browniennes infinie-dimensionelles,''
 Prob. Th. Rel. Fields {\bf 104}, 147--179 (1996).
\\

\bibitem[${ }^{3}$]{DerRoe05}
D. Dereudre and S. Roelly, 
``Propagation of Gibbsianness for infinite-dimensional gradient Brownian diffusions,'' 
J.Stat.Phys. {\bf 121}, 511--551 (2005).
\\
\bibitem[${ }^{4}$]{Deu87}
J.D. Deuschel, 
``Infinite-dimensional diffusion process as Gibbs measures on 
$C[0,1]^{\mathbb{Z}^d}$,'' 
Prob. Th. Rel. Fields {\bf 76}, 325--340 (1987).
\\

\bibitem[${ }^{5}$]{Dun79}
F. Dunlop, 
``Zeros of the partition function and Gaussian inequalities for the plane rotator model,''
J. Stat. Phys. {\bf 21}, 561-572 (1979).\\

\bibitem[${ }^{6}$]{Dun79a}
F. Dunlop, 
``Analyticity of the pressure for Heisenberg and plane rotator models,''
Comm. Math. Phys. {\bf 69}, 81-88 (1979). \\

\bibitem[${ }^{7}$]{vEntFerHolRed02}
A.C.D. van Enter, R. Fern\'andez, F. den Hollander and F.Redig, 
``Possible loss and recovery of Gibbsianness during the stochastic evolution of Gibbs measures,'' Comm. Math. Phys. {\bf 226}, 101--130 (2002).
\\

\bibitem[${ }^{8}$]{EFS93}
A.C.D. van Enter, R. Fern\'andez and A.D. Sokal, 
``Regularity Properties and Pathologies of Position-Space
Renormalization-Group Transformations: Scope and limitations of
Gibbsian Theory,'' 
J. Stat. Phys. {\bf 72}, 879-1169 (1993).\\


\bibitem[${ }^{9}$]{vEntRus07}
A.C.D. van Enter and W.M. Ruszel,
``Gibbsianness versus Non-Gibbsianness of time-evolved planar rotor models,''
arXiv 0711.3621, mp-arc 07-295 (2007), Stoch. Proc. and Appl, to appear.\\

\bibitem[${ }^{10}$]{FerPfi97}
R.Fernandez and C.-E. Pfister, 
``Global specifications and quasilocality of projections of Gibbs measures,''
Ann. of Prob. {\bf 25}, 1284--1315 (1997).
\\

\bibitem[${ }^{11}$]{FILS78}
J. Fr\"ohlich, R.B. Israel, E.H.Lieb and B.Simon, 
``Phase Transitions and Reflection positivity I,''
Comm. Math.Phys. {\bf 62}, 1-34 (1978).


\bibitem[${ }^{12}$]{Geo81}
H.-O. Georgii, 
``Percolation of low energy clusters and discrete symmetry breaking in classical spin systems,''
Comm.Math.Phys. {\bf 81}, 455--473 (1981).
\\

\bibitem[${ }^{13}$]{Geo88}
H.O. Georgii, 
``Gibbs measures and phase transitions,'' 
Berlin, W. de Gruyter, (1988).
\\
\bibitem[${ }^{14}$]{KueOpo07}
C. K\"ulske and A.Opoku, 
``The posterior metric and goodness of Gibbsianness of tranforms
of possibly continuous spin models,'' 
arXiv: 0711.3764 (2007), El.J.Prob., to appear.
\\
\bibitem[${ }^{15}$]{KueRed06}
C. K\"ulske and F.Redig, 
``Loss without recovery of Gibbsianness during diffusion of continuous spins,''
 Prob.Theor. Rel. Fields {\bf 135}, 428--456 (2006).
\\
\bibitem[${ }^{16}$]{Koz74}
O.K. Kozlov, 
``Gibbs description of a system of random variables,''
Probl. Info. Trans. {\bf 10}, 258--265 (1974).


\bibitem[${ }^{17}$]{LeNRed02}
A. Le Ny and F. Redig, 
``Short time conservation of Gibbsianness under local stochastic evolutions,'' 
J. Stat. Phys. {\bf 109}, 1073--1090 (2002).
\\


\bibitem[${ }^{18}$]{LieSok81}
E.H. Lieb and A. Sokal, 
``A general Lee-Yang theorem for one-component and multicomponent ferromagnets,'' Comm. Math. Phys. {\bf 80}, 153-179 (1981).

\bibitem[${ }^{19}$]{OlPe07}
M.J. de Oliveira and A. Petri, 
``Temperature in out-of-equilibrium lattice gas,'' 
Int. J. Mod. Phys. C, {\bf 17}, 1703-1715 (2007).

\bibitem[${ }^{20}$]{Ros97}
S.Rosenberg,
``The Laplacian on a Riemannian Manifold: An Introduction to Analysis on Manifolds,''
Cambridge University Press (1997).\\


\bibitem[${ }^{21}$]{Wehr2006}
 J.Wehr, A. Niederberger, L. Sanchez-Palencia and M. Lewenstein, 
``Disorder versus the Mermin-Wagner-Hohenberg effect: From classical spin systems to ultracold atomic gases,''
Phys. Rev. Vol B {\bf 74}, 224448 (2006).

\end{thebibliography}
\end{document}